\documentclass[twocolumn, tighten, times]{aastex63}
\usepackage[utf8]{inputenc}
\usepackage{bm}
\usepackage{tabularx}
\usepackage{braket}
\usepackage{multirow}
\usepackage{xcolor}
\usepackage{pifont}
\usepackage{enumitem}
\usepackage{amsmath}
\usepackage{soul}
\usepackage{diagbox}

\accepted{for publication in {\it The Astrophysical Journal}}
\shortauthors{Zhou, Hu \& Fang}

\graphicspath{{./}{figures/}}

\begin{document}

\title{A Distance-Deviation Consistency and Model-Independent Method to Test the Cosmic Distance-Duality Relation}
\author{Chichun zhou}
\author[0000-0002-4797-4107]{Jian Hu}
\email{hujian@dali.edu.cn}
\author{Maocai Li}
\author{Xi Zhang}
\affiliation{School of Engineering, Dali University, Dali 671003, China}
\author{Guanwen Fang}
\email{wen@mail.ustc.edu.cn}
\affiliation{School of Mathematics and Physics, Anqing Normal University, Anqing 246011, China}
\begin{abstract}
A distance-deviation consistency and model-independent method to test the cosmic distance duality relation (CDDR) is provided. The method is worth attention on two aspects: firstly, a distance-deviation consistency method is used to pair subsamples: instead of pairing subsamples with redshift deviation smaller than a value, say $\left\vert \Delta z\right\vert <0.005$. The redshift deviation between subsamples decreases with the redshift to ensure the distance deviation stays the same. The method selects more subsamples at high redshift, up to $z=2.16$, and provides 120 subsample pairs. Secondly, the model-independent method involves the latest data set of $1048$ type Ia supernovae (SNe Ia) and  $205$ strong gravitational lensing systems (SGLS), which are used to obtain the luminosity distances $D_L$ and the ratio of angular diameter distance $D_A$ respectively. With the model-independent method, parameters of the CDDR, the SNe Ia light-curve, and the SGLS are fitted simultaneously. The result shows that $\eta = 0.047^{+0.190}_{-0.151}$ and CDDR is validated at 1$\sigma$ confidence level for the form $\frac{{{D_L}}}{{{D_A}}}{(1 + z)^{ - 2}} =1+ \eta z$.
\end{abstract}


\section{Introduction} \label{sec:intro}
The cosmic distance duality relation (CDDR), also called Etherington's reciprocity relation \citep{eth33}, plays an important role in modern cosmology, especially in galaxy observations \citep{cmsl07,Manl10}, cosmic microwave background (CMB) radiation observations \citep{Kom11}, and the gravitational lensing \citep{ell07}. The CDDR reads
\begin{equation}\label{1}
\frac{{{D_L}}}{{{D_A}}}{(1 + z)^{ - 2}} = 1,
\end{equation}
where $D_L$ is the luminosity distance, $D_A$ is the angular diameter distance, and $z$ is the redshift. The CDDR is valid for all cosmological models based on Riemannian geometry. The basis of this relation is that the number of photons is conservative and photons travel along the null geodesic in a Riemannian space-time \citep{ell07}.

The validity of the CDDR is explored widely for the past decades, because any deviation of CDDR may trigger new physics. \citet{uza04} investigates the possible deviation from the CDDR by analysing the measurements of SZE and X-ray emission data of galaxy clusters and reports that the parameter $\eta = 0.89^{+0.04}_ {-0.03}$ and is at 1$\sigma$ confidence level. \citet{hlr11} takes more parametrized forms of $\eta$ and found no departure from the CDDR. \citet{njj11,baku04, hol10,hlr11,cali11,mzz12}used $D_L$ directly from type Ia supernovae (SNe Ia) to test the CDDR. \citet{hu18} and \citet{mel18} used the $D_L$ form SNe Ia with the $R_h=ct$ cosmology model to test the CDDR and compared the model with the $\Lambda$CDM model. \cite{baku04} found a 2$\sigma$ violation of the CDDR using the luminosity distances $D_L$  from SNe Ia and the angular diameter distance $D_A$ from FRIIb radio galaxies.  \cite{Rasa16} used CMB anisotropy to test the CDDR. The CDDR is also important in studying cosmic opacity \citep{lvxi16,hyw17}.

To test the CDDR, many $D_L$ and $D_A$ pairs at the same redshift $z$ need to be provided simultaneously. In principle, the two distances should neither be correlated nor based on any cosmology models. That is, a model-independent method and a quality and quantity collection of sample pairs are important. Conventionally, in determining $D_L$, the method of Standard Candles (e.g. SNe Ia, GRB \citep{wfy06,wfy07,wfy08,wfy09a,wfy09b,wfy15,wfy18,wfy19}) plays a prominent part. However, the method of Standard Candles is model-dependent, i.e., a special cosmology model is used in calibrating the light-curve parameters. For example, \citet{suz12} used the cold dark matter (CDM), wCDM, and owCDM models to fit the parameters of Union2.1 SNe Ia and to constrain the cosmology parameters. In determining $D_A$, the method of using the Sunyaev–Zel'dovich effect (SZE) and X-ray observations \citep{cafu78,suze72,bon06} is important in finding $D_A$ from galaxy clusters. $D_A$ can also be obtained from ultra-compact radio sources \citep{lili18} and baryon acoustic oscillations (BAOs) \citep{wu15}.

There are model-independent methods. \citet{lia16} introduced a new method to test the CDDR based on strong gravitational lensing systems and SNe Ia. In their work, they constrain  $\eta$, the parameter of the SNe Ia light-curve, and the parameter of the SGLS simultaneously. \citet{Rasa16} uses the temperature-redshift relation of CMB to test the CDDR, in their work, a flat FRW universe is assumed. \citet{rua18} use a similar model-independent method with the SGLS, the SNe Ia, and the HII galaxy Hubble diagram to test the CDDR. To avoid the effect of the cosmic opacity, \citet{lia19} uses the $D_L$ from the gravitational wave signals and the ratio of $D_A$ from the SGLS (for details see below) with a model-independent method, which is proposed in \citep{lia16}. These model independent methods show that the CDDR is valid in the given redshift range, say, $z<1.0$.

In this paper, we provide a distance-deviation consistency and model-independent method to test the CDDR. The distance-deviation consistency method pairs subsamples with redshift deviation decrease with the redshift to ensure the distance deviation stays the same. It is because the distance grows nonlinearly with the redshift, the larger the redshift is, the smaller the redshift deviation is between two sources with the same distance deviation. The latest data set of SNe Ia with $1048$ samples and strong gravitational lensing system (SGLS) with $205$ samples are involved and the distance-deviation consistency method enables us to take full advantage of the data: the method selects more subsamples at high redshift, up to $z=2.16$, and provide 121 subsample pairs. With the model-independent method, parameters of the CDDR, the SNe Ia light-curve, and the SGLS are fitted simultaneously. 

This paper is organized as follows. In Sec.2, we introduce the latest data set of the SNe Ia with $1048$ samples and the SGLS with $205$ samples. In Sec.3, we explain a distance-deviation consistency method to pair subsamples and describe the method of statistical analysis. The numerical results are shown. Conclusions and discussions are given in Sec.4.
\section{Data}\label{sec:Data}
In this section, we describe two sets of data suitable for testing the CDDR, one based on the redshift and the light-curve of the SNe Ia, from which we can obtain $D_L(z)$, the other base on the observational velocity dispersion of the lens galaxy, redshifts of the strong gravitational lensing system (SGLS), from which we can obtain the ratio of $D_A(z)$.

\subsection{The Pantheon SNe Ia Sample}\label{sec:The Pantheon SNe Ia Sample}
In this section, we introduce the contents of the Pantheon sample \citep{sco18}. The Pantheon sample consisting of a total of 1048 SNe Ia in the range of $0.01<z<2.3$ is constructed by a subset include 279 SNe Ia ($0.03<z<0.68$) from the Pan-STARRS1 (PS1) Medium Deep Survey, the Sloan Digital Sky Survey (SDSS), the SNLS, the various low-z, and Hubble Space Telescope (HST) samples.

On one hand, the luminosity distances $D_L(z)$ can be determined accurately by multiple light-curve fitters (e.g., \citep{jha07, guy10, bur11, man11}). From the phenomenological point of view, the distance modulus, $\mu$, of an SNe Ia can be extracted from its light curve. In a modified version of the Tripp formula \citep{trip98}, the SALT2 light-curve fit parameters are transformed into distances:
\begin{equation}\label{2}
\mu= m_B - M + \alpha x_1 - \beta c + \Delta_M + \Delta_B,
\end{equation}
where $m_B$ is the apparent magnitude, $M$ is the absolute B-band magnitude of a fiducial SNe Ia with $x_1=0$ and $c=0$, $\Delta_M$ and $\Delta_B$ are distance corrections based on the mass of the host galaxy of the SNe Ia and predicted biases from simulations respectively. $\alpha$ and $\beta$ are light-curve parameters of relations between the luminosity and the stretch and between the luminosity and the color respectively. Moreover, $\Delta_M$ in the equation (\ref{2}) can be written in the form \citet{sco18}
\begin{equation}\label{4}
\Delta_M = \gamma[1 + e^{(-(m-m_{step})/\tau)}]^{-1},
\end{equation}
where  $\gamma$, $m_{step}$, and $\tau$ are coefficients to be determined \citep{sco18}.
On the other hand, one assumes that the SNe Ia with identical color, shape, and galactic environment have on average the same intrinsic luminosity for all redshifts \citep{bet14}. According to the definition of the distance modulus, it can be written as
\begin{equation}\label{3}
\mu = 5 \log(\frac{D_L}{\rm Mpc})+25.
\end{equation}
By using equations (\ref{2}) and (\ref{3}), we can obtain the luminosity distances $D_L(z)$ for the Pantheon SNe Ia Sample.

\subsection{The strong gravitational lensing system Sample}\label{sec:SGLS}
For the new SGLS sample, we used from \citet{ama19}, which contains 205 SGLS. This sample constituted from some survey projects, (e.g. the SLACS, the CASTLES survey, the BELLS, the LSD ...).
In an SGLS, the light is bent by massive bodies (e.g., galaxy, galaxy cluster) which is predicted by the general theory of relativity. The SGLS is a powerful astrophysical tool to explore the universe and galaxy and has been rapidly developed in recent years, especially like dark energy \citep{bie06, bie10, cao12, cao15, jul10, mag15,mag18}, the CDDR \citep{lia16, lia19}, the cosmic acceleration \citep{tu19}, calibrating the standard candles \citep{wen19}, and cosmological models' comparison\citep{mel15,lea18,tu19}.
In an SGLS, a single galaxy acting as the lens, the Einstein radius depends on three parameters: depends on the angular distance to the source and between the lens and the source, and the mass distribution within the lensing galaxy. A singular isothermal sphere (SIS) model \citep{rat99} is used to describe the lens galaxy’s mass distribution. The ratio of the angular diameter distances between lens and source and between observer and source can be obtained from a special physical model (e.g. SIS model). Because these distances depend on the cosmological metric, the ratio can be used to constrain cosmological parameters.

In an SIS model of the SGLS, the distance ratio $R^A(z_l,z_s)$ ~~($D^A_ls/D^A_s$) is related to observables in the following way \citep{bie10},
\begin{equation}\label{5}
R^A(z_l,z_s)=\frac{c^2\theta_E}{4\pi\sigma^2_{SIS}},
\end{equation}
where $c$ is the speed of light, $\theta_E$ is the Einstein radius, and $\sigma_{SIS}$ is the velocity dispersion of the stellar in the periphery of the lens galaxy due to the lens mass distribution in the SIS model. In general, $\sigma_{SIS}$ not equals to the observed stellar velocity dispersion $\sigma_0$ \citep{whi96}. To express the difference, researchers use a phenomenological free parameter $f_e$  defined by the relation $\sigma_{SIS} = f_e\sigma_0$ \citep{koc92, ofe03, cao12}, where $(0.8)^{1/2}  <f_e< (1.2)^{1/2}$. In this case, the systematic error is caused by $\sigma_0$ as $\sigma_{SIS}$, the deviation of the SIS model, the effects of secondary lenses (nearby galaxies), the line of sight contamination \citep{ofe03}, etc. The uncertainty of equation (\ref{5}) can be written by \citep{lia16}
\begin{equation}\label{6}
\sigma_{R^A}(z_l,z_s)=R^A(z_l,z_s)\sqrt{(4\delta_{\sigma_{sis}})^2+(\delta_{\theta_E})^2}.
\end{equation}
In equation(\ref{6}), $\delta_{\sigma_{sis}}$ and $\delta_{\theta_E}$ are the fractional uncertainty of the 
$\sigma_{sis}$ and $\theta_E$, respectively. To test the CDDR, the left term of equation (ref{5}), $R^A(z_l,z_s)$, should be expressed as the ratio of luminosity distance, $D^A_{ls}/D^A_s$. We transform $R^A(z_l,z_s)$ into the ratio of Comoving distance($D_C$) or dimensionless distance $H_0D_C/c$. In a flat space, the dimensionless distance satisfies
\begin{equation}\label{7}
d(z_l,z_s)=d(z_s)-d(z_l).
\end{equation}
By using the equation,
\begin{equation}\label{8}
D_A(z)=\frac{D_C(z)}{1+z}=\frac{H_0d(z)}{c(1+z)},
\end{equation}
$R^A(z_l,z_s)$ can be written as
\begin{equation}\label{9}
R^A(z_l,z_s)=1-\frac{(1+z_l)D_A(z_l)}{(1+z_s)D_A(z_s)}.
\end{equation}

In a non-flat space, the expression of $R^A(z_l,z_s)$ is more complicated, one can refer to \citet{2015Syksy}. Fortunately, most cosmological tests today support a flat cosmic \citep{planck}, in this work, we test the CDD relation in the case where we assume that space-time is flat.

To test the CDDR, equation (\ref{1}) is rewritten by the parametrization of the deviation:
\begin{equation}\label{11}
\frac{{{D_L}}}{{{D_A}}}{(1 + z)^{ - 2}} =1+ \eta z.
\end{equation}
Combining equations (\ref{9}) and (\ref{11}), $R^A(z_l,z_s)$ can be written as
\begin{equation}\label{12}
R^A_0(z_l,z_s)=1-\frac{(1+\eta z_s)(1+z_s)d_L(z_l)}{(1+\eta z_l)(1+z_l)d_L(z_s)}.
\end{equation}
By using equation (\ref{3}), the part $\frac{d_L(z_l)}{d_L(z_s)}$ of equation (\ref{12}) can be rewritten as
\begin{equation}\label{13}
\begin{split}
\lg[\frac{d_L(z_l)}{d_L(z_s)}]=&0.2\{m_{B}(z_l)-m_{B}(z_s)+ \alpha[x(z_l)-x(z_s)]-\\
&\beta[c(z_l)-c(z_s)]+\Delta_M(z_l)-\Delta_M(z_s)\},
\end{split}
\end{equation}
where, the absolute magnitude of SNe Ia subsample is offset, $z_l$ is the redshift of the lens, and $z_s$ is the redshift of the sources.
\section{Method and results} \label{sec:Method and results}
In this section, we introduce a distance-deviation consistency method of data selection and show the result.

\subsection{Method of data selection: a distance-deviation consistency method}\label{sec:method of data selection}

To test the CDDR, $D_L$ and $D_A$ at the same redshift $z$ need to be provided simultaneously. However, redshifts of subsamples from the SGLS and the SNe Ia are different. To take full advantage of the data, one needs to pair the subsamples efficiently. In this section, we provide a distance-deviation consistency method to pair subsamples, which outperforms the conventional method.

The redshift-difference of subsample-pairs in this work is not fixed, and it decreases with the redshifts to ensure the distance deviation of the sources stays the same. The relation between $\Delta z$ and coordinate distance with a cosmology model reads
\begin{equation}\label{14}
\Delta{d_c^{model}(z)}=d_c^{model}(z+\Delta z^{model})-d_c^{model}(z),
\end{equation}
where $R_h=ct$ and flat $\Lambda$CDM model with $\Omega_m=0.31$\citep{planck} are used. By setting $\Delta d_c/d_c$ equals 5\% and combining equation(\ref{14}), $\Delta{z}(z)$ can be calculated:
\begin{equation}\label{15}
\Delta{z}(z)=min\{\Delta{z}^{\Lambda CDM}(z),\Delta{z}^{R_h=ct}(z)\},
\end{equation}
where the distance formulas of the two cosmology model, $\Lambda$CDM and $R_h=ct$ are used. The $R_h = ct$ cosmological model was proposed in \citet{mel07}. In the $R_h = ct$ universe, the luminosity distance $D_L$ can be written by:
\begin{equation}\label{151}
D_L^{R_h=ct}(z)=\frac{c}{H_0}(1+z)\ln(1+z)
\end{equation}
This model is a Friedmann–Robertson–Walker (FRW) cosmological model, which is obeying the cosmological principle and Weyl's postulate\citep{mel07,meshl12}. In $R_h=ct$ universe that space expands at a constant rate, rather than an accelerating rate. Theoretically, there are some controversies with this model, which focus on the zero active mass condition $\rho$$+3p = 0$(for more details see \citet{mel16,klh16,mel17}). In terms of the fitting of observational data, the model performs relatively well. The originator of this model himself and his collaborators have done extensive work comparing it with the standard model using many different types of observations, and they have found that the $R_h=ct$ model is better than the Standard Model(e.g.\citet{mel13,melm18,fame17,mel14,jjw15,mel19}). Additional work by others also illustrates this point\citep{yuwa14,yuwa15}. There is also a lot of work against this model(e.g. \citet{tut16,shaf15}) Thus, despite the theoretical controversy, this model has gained some support for the data, and we can use this model jointly with the standard model to select the data.

For a given SGLS subsample, the redshift-dimensionless distance relation, equation(\ref{15}), is used to acquire a suitable redshift deviation interval. Then, the subsamples of the SNe Ia with redshifts within the interval are selected as the candidate. Finally, the subsample with the smallest redshift-deviation is selected.

Our method outperforms the conventional method in two aspects:
    (1) information of subsamples at high-redshifts is conserved. To pair subsamples with a slight difference of redshift is a simple and commonly used method. For example, $\Delta z=0.005$ \citep{hol10,hol12,hba16,li11, njj11},  $\Delta z=0.006$ \citep{ghj12}, and $\Delta z=0.003$ \citep{lia19}. Gaussian Process (GP) reconstruction is also a usable method \citep{zha14, rua18}. The linear interpolation method is also taken by some researchers \citep{liang13, hu18}. These methods reduce the systematic error to an extent but do not consider the distance-deviation consistency. The relation between the distance and the redshift is non-linear: the same redshift deviation at the higher redshift has a smaller distance deviation. For example, some researchers set the $\Delta z=0.005$, the uncertainty of dimensionless distance is 5\% at $z\sim0.1$, the minimum redshift of the subsample we select, but small than 1\% at $z\sim1$ for selecting data with a general cosmology model, so the selecting uncertainty is not the consistency of distance-deviation. The selecting uncertainty of their methods are all ignoring, and if they were taking them into account in their fitting, they had to introduce a cosmological model so that their methods were no longer model-independent. In our approach, although two cosmological models are introduced, they are used to jointly pick the data, breaking the dependence on a single cosmological model when picking the data, and are not introduced into the $\chi^2$ function, in other words, the final parameter fit results are independent of the cosmological model. \citet{cao17}used a similar method for selecting the data with a $\Lambda$CDM model to fitting the parameters of the ultra-compact radio quasars. At high redshift, in general, subsamples are sparse and matching pairs of subsamples is difficult if a fixed $\Delta z$ is used. The distance deviation consistency method selects more subsamples at high-redshifts.

\begin{figure}
  \centering
  \includegraphics[scale=0.38]{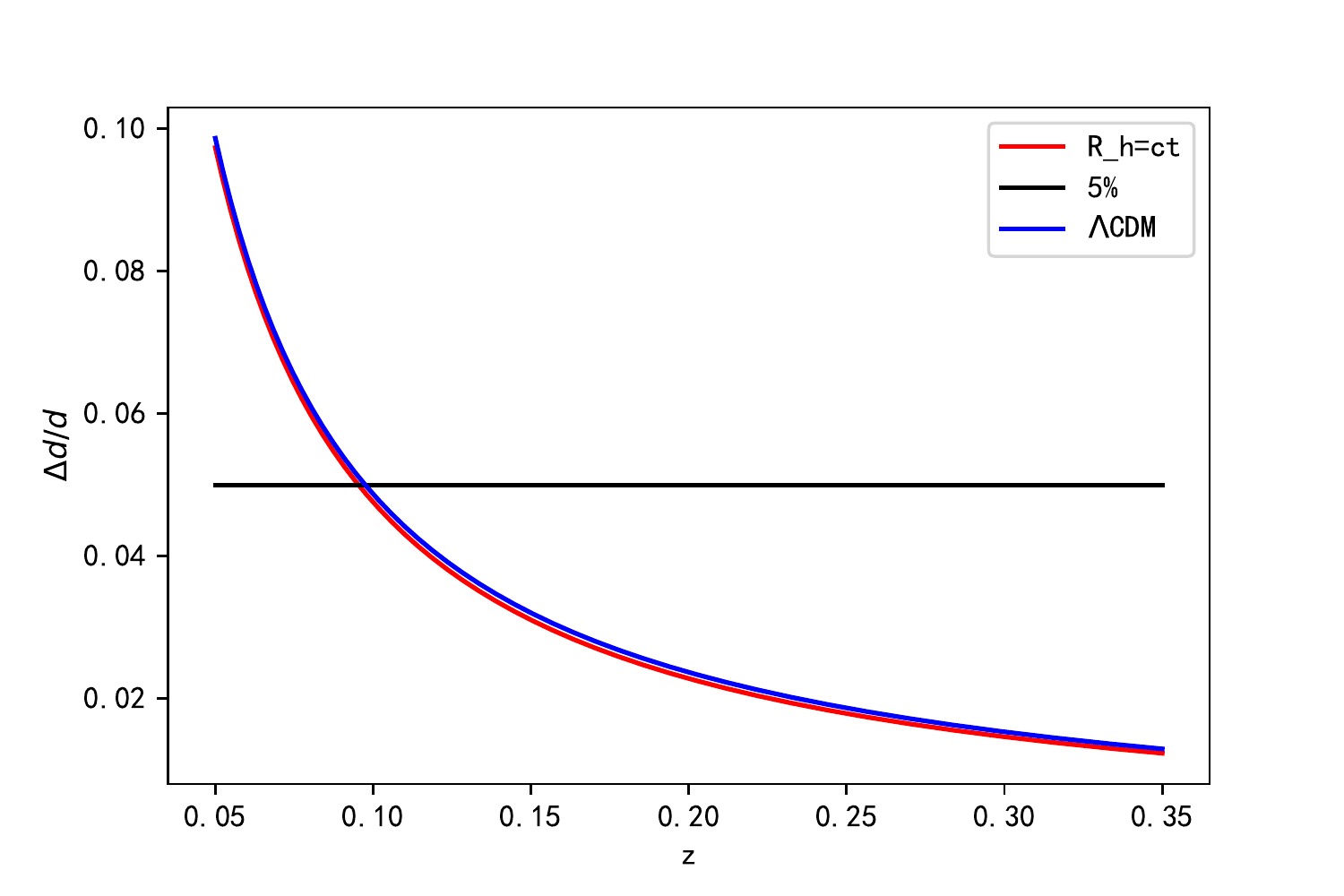}\\
  \caption{The relation between the $\Delta d/d$ and $z$,the red line and the blue line indicate the relationship between the relative error of $d$ and the redshift in $R_h=ct$ model  and $\Lambda CDM$ model, respectively. The black line indicates the scheme we used.}
  \label{fig01}
\end{figure}

    (2) More subsample pairs are selected. For example, in the work of \citet{lia16}, they gain only 60 pairs of samples to testing the CDDR. According to figure(\ref{fig01}), at z=0.11, the most previously used $\Delta z=0.005$ and the curve of $\Delta d/d$ in this method are intersecting. So,  the $\Delta d/d=5\%$ is the maximum allowable uncertainty of dimensionless distance. In other words, if the error of selecting exceeds 5\%, the statistical error of our results must be higher than that of previous work. If the error is significantly lower than 5\%, the number of data pairs we choose will not be significantly improved. With the method of fixed $\Delta d/d=5\%$, the number of the pairs can reach 68. The data utilization increases by 13.3\%. In this work, we have obtained 120 pairs of samples that redshift from 0.11 to 2.16.

\begin{figure}
  \centering
  \includegraphics[scale=0.38]{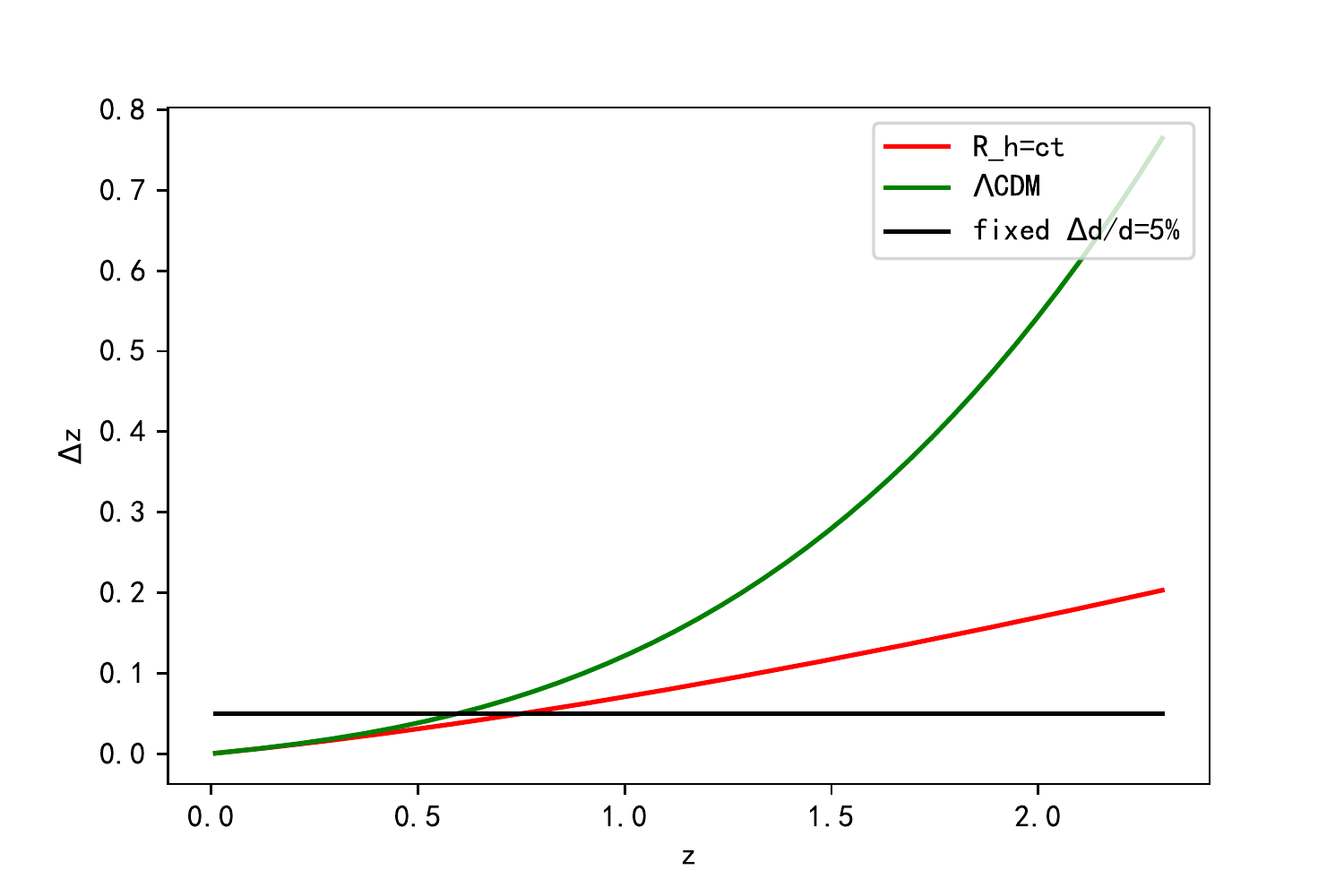}\\
  \caption{The relation between the $\Delta_z$ and $z$}
  \label{fig02}
\end{figure}

A comparison between our method with a fixed $\Delta d/d$ and the conventional method with a fixed $\Delta z$ are shown in figures (\ref{fig01}) and (\ref{fig02}). In figure (\ref{fig01}), the deviation, $\Delta d/d$, falls rapidly with the increase of redshift with $\Delta z$  fixed.  The main advantage of a fixed $\Delta d/d$  is shown in figure (\ref{fig02}). To conclude, on the one hand, our method maintains more information about subsamples at high-redshifts. On the other hand, our method selects more subsample-pairs thus reduces the systematic error.

\subsection{Method of statistical analysis}\label{sec:method of statistical analysis}
To determine the parameters, we minimize $\chi^2$ function. By using equations (\ref{5}) and (\ref{12}), $\chi^2$ function can be written as
\begin{equation}\label{16}
\chi^2=\sum^{120}_{1}(\frac{(R^A(z_l,z_s)-R^A_0(z_l,z_s))^2}{\sigma^2_{R^A(z_l,z_s)}+\sigma^2_{R^A_0(z_l,z_s)}+\sigma^2_{sel}}),
\end{equation}
where $\sigma_{R^A(z_l,z_s)}$ is the uncertainty of the SGLS with SIS model, $\sigma_{R^A_0(z_l,z_s)}$ is the error caused by the uncertainty of the distance modulus of SNe Ia, which is related to the uncertainty of observed data (e.g. $m_B$), and $\sigma_{sel}$ is the uncertainty of data selection, which is related to artificial selection. By using equation(\ref{12}), the uncertainty of data selection can be written as
\begin{equation}\label{17}
\sigma_{sel}=\frac{(1+\eta z_s)(1+z_s)d_L(z_l)}{(1+\eta z_l)(1+z_l)d_L(z_s)}*\sqrt{(\frac{\Delta d_l}{d_l})^2+(\frac{\Delta d_s}{d_s})^2},
\end{equation}
where $\frac{\Delta d_l}{d_l}=\frac{\Delta d_s}{d_s}=\frac{\Delta d}{d}=5\%$. 

We use the Markov chainMonte Carlo (MCMC) method to constrain the parameters in equation(\ref{17}). EMCEE \citep{fore13} python package is used. In order to execute the MCMC process, we need to provide the prior values first. In  \citet{sco18}, the best value and the 1$\sigma$ confidence level of the parameters are shown. But in our work, we take an SIS model of SGLS to calibrating these parameters, which will different. So, we take the prior interval that completely covering the value range of the value from \citet{sco18}. The prior probability for parameters $P(\alpha,\beta,f_e,\eta,\gamma, m_{step},\tau)$ is the product of  prior probability of each parameter. The prior probability is assumed to be uniform distributions: $P(\alpha) = U [-0.2, 0.2]$, $P(\beta) = U [2, 6]$, $P(f_e)=U [0.5, 1.5]$, $P(\eta) = U [-0.5,1.5]$, $P(\gamma) = U [0,0.3]$, $P(M_{step}) = U [5,15]$, $P(\tau) = U [0.001,1]$. In Pantheon samples, the errors include both statistic and systematic deviation. The systematic error relatives to all data point and appears as a huge covariance matrix. In this work,  part of SNe Ia are selected, only the statistic error is considered.

\subsection{Results}
\begin{figure*}
  \centering
  \includegraphics[scale=0.38]{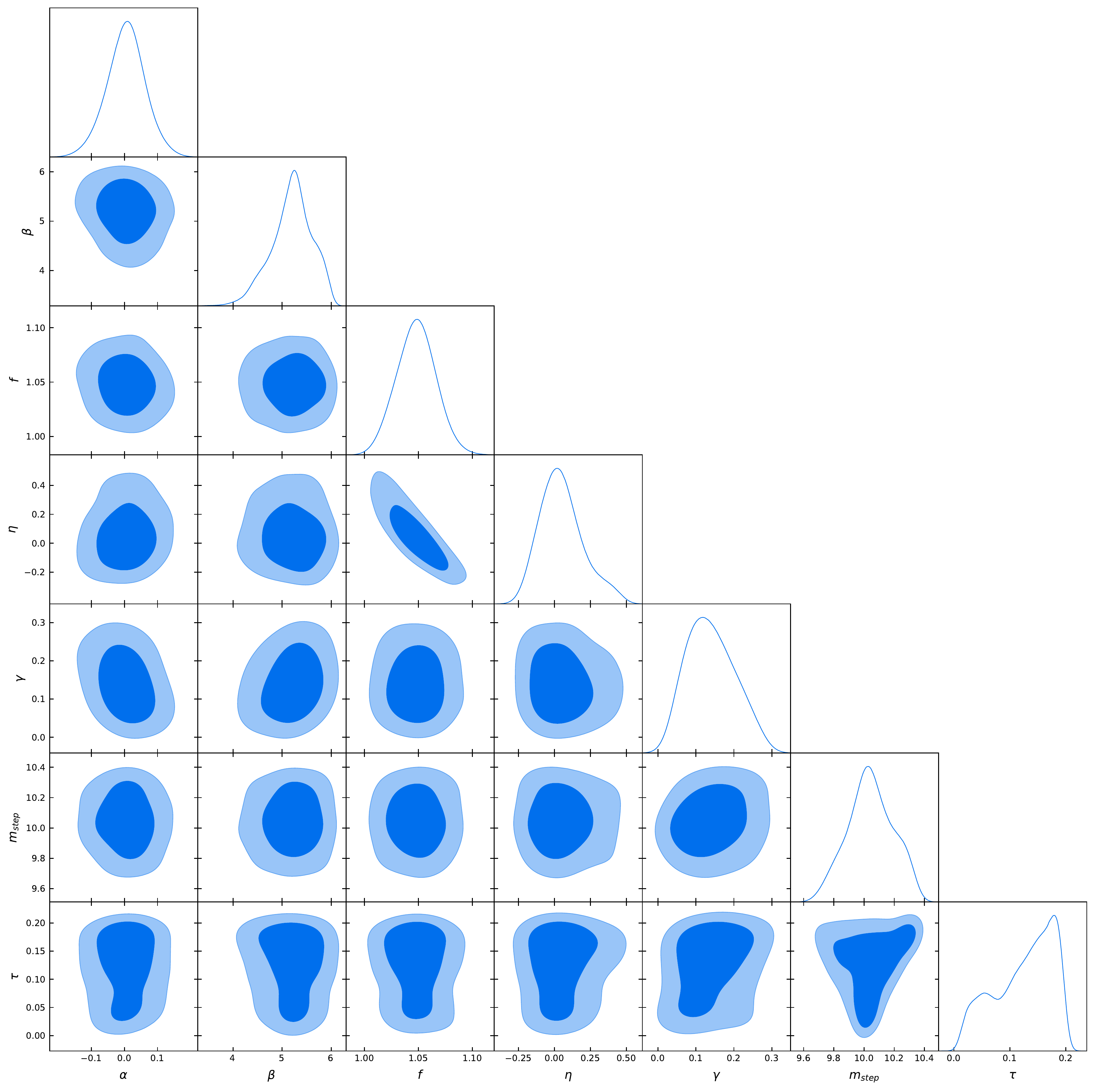}\\
  \caption{The 2D regionsand 1D marginalized distributions with 1$\sigma$ and 2$\sigma$ contours for the parameters $\alpha$, $\beta$, $f_e$, $\gamma$, $m_{step}$,$\tau$ and $\eta$ using Pantheon sample and SGLS sample.}
  \label{fig03}
\end{figure*}

The result is shown in figure (\ref{fig03}) and table (\ref{tab1}). Triangle contours are plotted by using the open-source python package “Getdist”. One can see from figure (\ref{fig03}) that the best-fitted center value is $\eta = 0.047^{+0.190}_{-0.151}$, which is at 1$\sigma$ confidence level. The result indicates that the CDDR is in agreement with the observations and there are no signs of violation in light of SN Ia and SL data.

\begin{table}
\caption{Constraints on the coefficients of light-curve parameters
and $\eta$ at the 1$\sigma$ confidence levels.} \scalebox{0.9}{
\begin{tabular}{|c|c|c|}
\hline
         parameters  &             value    \\
\hline
        $\alpha$  &  $0.001^{+0.061}_{-0.061}$ \\
\hline
        $\beta$  &        $5.283^{+0.417}_{-0.464}$      \\
\hline
        $f_e$  &        $1.046^{+0.020}_{-0.019}$ \\
\hline
        $\eta$ &        $0.047^{+0.190}_{-0.151}$           \\
\hline
        $\gamma$ &        $0.141^{+0.080}_{-0.070}$       \\
\hline
        $m_{step}$ &         $10.055^{+0.177}_{-0.148}$  \\
\hline
        $\tau$ &         $0.134^{+0.048}_{-0.073}$   \\
\hline
        $\chi^2$ &       117.430  \\
\hline
        $\chi^2/{d.o.f }$ &   117.430/113  \\
\hline
\end{tabular}}\label{tab1}
\end{table}

\section{DISCUSSIONS AND CONCLUSIONS} \label{sec:DISCUSSIONS AND CONCLUSIONS}
The validation of the CDDR is a crucial topic in cosmology. Any violation of the CDDR may generate a new theory of physics. In recent years, to compare the $D_L$ derived from SNe la and the $D_A$ measured using galaxy clusters is the common method to test the CDDR. To use this method, a specific cosmology model with some parameters (e.g. the matter density parameter $\Omega_m$, the cosmic equation of state, and the Hubble constant) must be assumed. Such results are hardly convincing.

In testing the CDDR, using a model-independent method is necessary. Moreover, to obtain a data sample contains a large number of $D_L$ and $D_A$ pairs are also important. However, the number of useful subsample pairs is limited by the observed data, and  pair subsamples with redshift-deviation smaller than a constant will lose the subsamples at high redshift which leads to systematic errors.

In this paper, we provide a distance-deviation consistency and a model-independent method to test the CDDR. By applying the distance-deviation consistency method on the latest data set of SNe Ia with $1048$ samples and strong gravitational lensing system (SGLS) with $205$ samples, we obtain the collection of subsample pairs not only contains more subsamples but also maintains the information of subsamples at high redshift, up to $z=2.16$. By applying a model-independent method: the SGLS model is used to replace the cosmology model in SNe Ia light-curve fitting, the result shows that $\eta = 0.047^{+0.190}_{-0.151}$ and CDDR is validated at 1$\sigma$ confidence level for the form $\frac{{{D_L}}}{{{D_A}}}{(1 + z)^{ - 2}} =1+ \eta z$.

\acknowledgments
We thank the anonymous referee for constructive comments.This work is supported by the Research Foundation for Advanced Talents(NO KY1916102940,KY1916102740), Dali university and the Special Foundation for Basic Research Plan (grants KY2013113740 and KY2013114440)

\bibliographystyle{aasjournal}

\end{document}